\documentclass[aps,prb,superscriptaddress,twocolumn]{revtex4}
\usepackage[dvips]{graphicx} \usepackage{amsmath}

\begin{document}


\title{Quantum size effects in Pb islands on
Cu(111): Electronic-structure calculations}

\author{E. Ogando}\email{eoa@we.lc.ehu.es}\affiliation{Elektrika eta
Elektronika Saila, Zientzia Fakultatea UPV-EHU 644 P.K., 48080 Bilbao, Spain}
\author{ N. Zabala}\affiliation{Elektrika eta Elektronika Saila, Zientzia
Fakultatea UPV-EHU 644 P.K., 48080 Bilbao, Spain} \affiliation{Donostia
International Physics Center (DIPC),  Paseo de Manuel Lardizabal 4,
20018 Donostia, Spain} \author{E.V. Chulkov}\affiliation{Donostia
International Physics Center (DIPC), Paseo de Manuel Lardizabal 4,
20018 Donostia, Spain} \affiliation{Materialen Fisika Saila, Kimika
Fakultatea, UPV-EHU 1072 P.K., 20080 Donostia, Spain} \author{M.J. Puska}
\affiliation{Laboratory of Physics, Helsinki University of Technology,
P.O. Box 1100, FIN-02015 HUT, Finland }

\date{\today}

\begin{abstract}
The appearance of "magic" heights of Pb islands grown on Cu(111) is studied
by self-consistent electronic structure calculations. 
The Cu(111) substrate is modeled with a one-dimensional pseudopotential 
reproducing the essential features,
i.e. the band gap and the work function, of the Cu band structure in 
the [111] direction. Pb islands are presented as stabilized jellium
overlayers. The experimental eigenenergies of the quantum well states confined 
in the Pb overlayer are well reproduced. The total energy oscillates as a continuous 
function of the overlayer thickness reflecting the electronic shell structure.
The energies for completed Pb monolayers show a modulated oscillatory pattern
reminiscent of the super-shell structure of clusters and nanowires.
The energy minima correlate remarkably well with the measured most probable heights
of Pb islands. The proper modeling of the substrate is crucial to set the 
quantitative agreement.

\end{abstract}

\pacs{73.21.Fg, 71.15.Nc, 71.15.Mb, 68.35,-p }


\maketitle


The confinement of valence electron states in low-dimensional systems has 
a strong influence on the size distributions of nanostructures grown or 
produced in experiments. In clusters of alkali metal atoms the confinement occurs in
all three dimensions and the structures corresponding to closed electronic shells
are the most stable and therefore the most abundant ones.
\cite{knight84} Alkali metal nanowires \cite{yanson2} and metallic overlayers on 
solid surfaces \cite{hinch,crommie,otero} exhibit two- and one-dimensional (1D)
confinement, respectively. In these systems the sinking of the 
bottom of a new subband below the Fermi level is accompanied with an increase 
in the total energy destabilizing the system. The detailed understanding
of the mechanisms controlling the growth of nanoobjects is of vital
importance when producing, and eventually manufacturing, highly-organized
atomic-scale structures in nanodevices.

In this work we focus on the 1D confinement in Pb islands on the Cu(111) surface. 
The growing of Pb occurs in the [111] direction.
Hinch {\em et al.} \cite{hinch} studied the island height distribution of Pb using 
the He atom scattering (HAS). Recently, Otero {\em et al.} \cite{otero} have determined 
the height distribution of Pb islands using the scanning tunneling microscope (STM) up 
to heights over 20 Pb monolayers (ML's). The lateral dimension of the islands is large, 
of the order of 500 \AA, justifying the 1D modeling of their energetics. The vertical 
confinement in the Pb overlayer is due to the potential barrier between Pb and the vacuum 
and the energy gap in the projection of the Cu bulk bands in the [111] direction.
A closed shell occurs periodically when the thickness $D$ of the Pb layer satisfies 
$D=n\lambda_F/2$, where $\lambda_F$ is the Fermi wavelength of Pb and $n$
is an integer. The effect of this shell structure on the physical properties
was recognized already in early calculations \cite{schulte} of unsupported jellium slabs and
the stability as a function of the thickness was discussed on the basis of 
density-functional theory (DFT) calculations already two decades ago. \cite{feibelman} 
Later jellium and pseudopotential calculations \cite{sarria,kiejna,saalfrank,materzanini,wei} 
have dealt with quantum size effects in unsupported  metallic slabs, while the effects of 
the substrate have not been explicitly considered.

We model the whole system consisting of the Pb overlayer and the Cu(111) substrate 
by using the stabilized jellium (SJ) model\cite{perdew} 
for the overlayer and a 1D pseudopotential for the substrate and obtain results 
with a quantitative predictive power. We calculate electronic structure using 
local-density approximation (LDA). The delocalized electron character of Pb at
the Fermi level justifies the use of the SJ model
with $r_s=2.3a_0$ to describe the Pb overlayer. The SJ model
gives the work function of 4.1 eV, which is close to the experimental value
of 4.0 eV for Pb. \cite{ashcroft} The relevant feature giving 
physical insight is that the SJ model allows us to simulate overlayers of 
any thickness. For the Cu(111) surface we have constructed a new 1D 
pseudopotential. We start from the 1D model potential by
Chulkov {\em et al.}, \cite{chulkov} cosinelike bulk part of which 
reproduces correctly the experimental 
energy gap of Cu in the (111) direction. The position of the Fermi level with
respect to the model potential determines the electron density.
We subtract then the interactions between valence electrons from the effective potential 
within the LDA and obtain a local 1D pseudopotential. The pseudopotential
is constructed to give the experimental Cu(111) work function
of 4.94 eV.\cite{chulkov} 
The self-consistent screening of this 1D pseudopotential within 
the LDA accounts correctly for the energy gap in the (111) projection, 
both in the width and in the position with respect to the Fermi level. 
The details of the construction are described in a forthcoming paper. \cite{ogando}

\begin{figure}[t]
\includegraphics[width=\columnwidth]{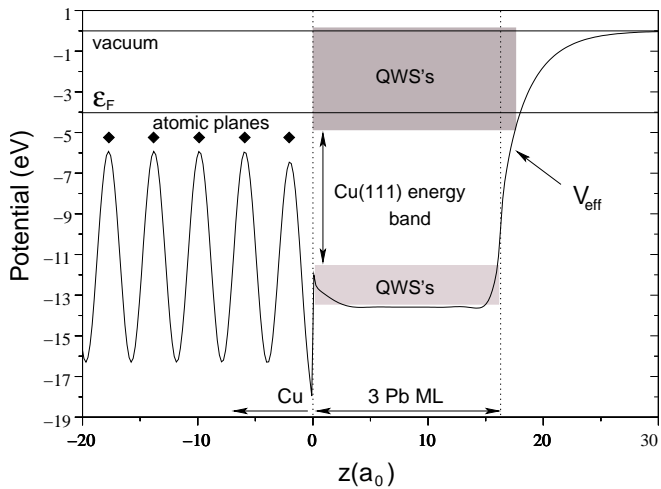}
\caption{\label{pot} Effective potential as a function of the distance 
to the Cu/Pb interface. 
The vertical dotted lines give the limits of the 3 ML Pb slab. 
The dark gray area corresponds to the Cu(111) energy gap and 
the light gray to the potential well at Pb due to its deeper effective 
potential compared with the average of Cu(111) potential.}
\end{figure}

In the numerical calculations a semi-infinite crystal is simulated by the slab 
geometry in which 25 layers of Cu are covered on both sides by Pb-SJ with the desired
thickness. In Fig. \ref{pot} we show the effective potential in the case of 3 Pb ML. 
One ML corresponds to the thickness $d=5.441a_0$. In Cu, the potential maxima 
due to the 1D-pseudopotential mimics the orthogonalization 
of the valence states to the core states. At the Pb/Cu interface
the potential is affected by the alignment of the Fermi levels of the
substrate and the overlayer accompanied by the charge transfer from Pb to Cu.
In the Pb layer the potential is a rather flat well and at the Pb-vacuum 
interface it forms a dipole-layer step. 
The dark gray region denotes the energy gap induced by the Cu potential 
and the light gray region gives the energy range between the potential in
Pb and the bottom of the Cu bulk band. These two regions will accommodate the
quantum-well states (QWS) localized perpendicular to the surface and mainly in 
the Pb overlayer. QWS's at the Cu band gap determine the energy shell structure.

\begin{figure}
\includegraphics[width=\columnwidth]{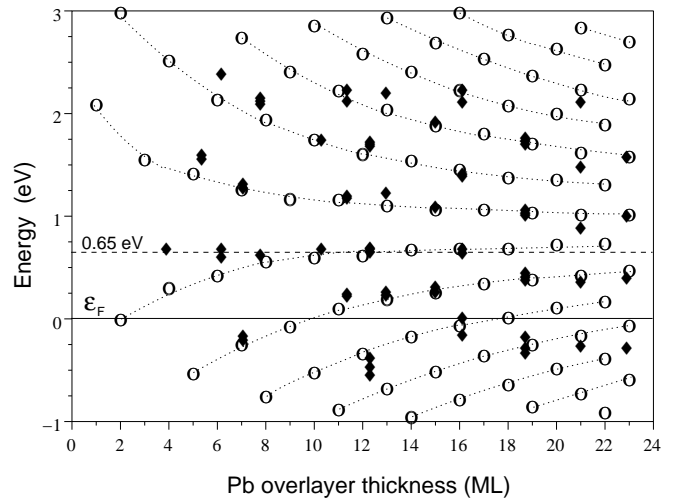}
\caption{\label{eigen} Eigenenergies of the QWS's as a function of
the Pb layer thickness.
Open circles and filled diamonds show the calculated and
measured \cite{otero2} values, respectively. The dotted lines are plotted as a 
guide to the eye.
The energies are given with respect to the Fermi
level of Pb/Cu(111) system.}
\end{figure}

The QWS eigenenergies (bottoms of two-dimensional paraboloid bands) are shown in Fig. \ref{eigen}
as a function of the number of Pb ML's completed. They are compared
with the results measured by STM. \cite{otero2}
The agreement is good for thick Pb coverages (thicker than 6 ML of Pb). 
In particular, both the theory 
and experiment give a QWS at $\approx 0.65$ eV for every even number of ML's
(no experimental data are available for 18, 20, and 22 ML).
For coverages less than 7 ML the correspondence is worse. This disagreement 
can reflect the compression of the Pb layers in both the vertical (perpendicular 
to the interface) and in horizontal (parallel to the interface) directions omitted in 
our simple model. \cite{materzanini,sarria} The effect of the compression is to push
up the eigenenergies the stronger the thinner the Pb overlayer is. Another important 
factor that can influence the QWS energies for thin coverages of Pb and which is not 
taken into account in our model is the interaction of 3d electrons of Cu with delocalized
electrons of Pb. Due to the very localized character of 3d electrons this interaction 
takes place mostly at the Cu-Pb interface and disappears at distances of a few ML's 
of Pb because of screening by Pb electrons. The overall agreement between the theoretical 
and measured eigenenergies for thick Pb overlayers gives 
confidence to our model when studying the behaviour of the total energy.

\begin{figure}[t]
\includegraphics[width=\columnwidth]{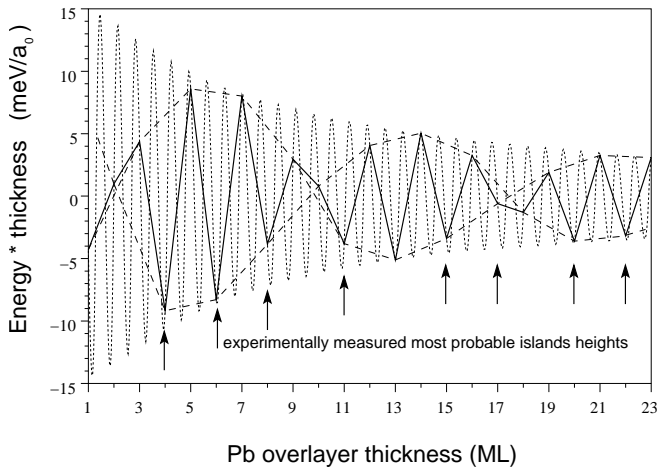}
\caption{\label{sigma} Oscillating part of the total energy 
per surface area of Pb/Cu(111) system multiplied by the Pb slab thickness. The dotted line 
is a function of the continuous Pb-SJ thickness. The solid line
connects the values corresponding to completed ML's.
The dashed lines guide the eye in order to show the supershell structure. 
The arrows denote the measured thicknesses of the most abundant
island heights. \cite{otero} }
\end{figure}

The total energy of the Pb/Cu(111) system 
can be written as the sum $E_{Cu} + E_{Pb} + E_{Cu-Pb} + E_{Pb-vac}$, where 
$E_{Cu}$ and $E_{Pb}$ are contributions calculated by using the
bulk energies corresponding to the 1D-Cu(111) pseudopotential and 
the Pb-SJ, respectively. The sum of the energies due to the Cu-Pb and Pb-vacuum interfaces,$E_{Cu-Pb}+E_{Pb-vac}$,  contains the oscillating part of the
total energy. In order to see the oscillations more clearly also at high
coverages Fig. \ref{sigma} shows the oscillating part of the energy multiplied 
by the Pb thickness. As a continuous function of the Pb thickness the
oscillations are regular and their wavelength is half the Fermi
wavelength, $\frac{\lambda_F}{2} =3.77 a_0$, of the Pb jellium.

The solid line in Fig. \ref{sigma} links the points corresponding to 
completed Pb ML's in the [111] direction. Because the interlayer spacing $d \simeq 3/2 \frac{\lambda_F}{2}$ an
even-odd staggering of the energy as function of the ML's
is obtained. However, the above relation is not exact and therefore 
the staggering amplitude diminishes regularly and the phase of
the staggering changes at the beats from that corresponding to minima at 
an even number of ML's to minima at an odd number of ML's or vice versa. The pattern 
resembles that of the supershell structures for atomic clusters
\cite{heer} and metallic nanowires. \cite{yanson1} Therefore
we will speak in the present context also about the supershell structure
although now the physical origin is different.

The arrows pointing upward in Fig. \ref{sigma} denote the most
abundant island heights measured by Otero {\em et al.} \cite{otero}
with the STM.
Actually, the measurements give the percentage of area covered 
with islands of a given height and the highlighted heights correspond
to maxima in this distribution.
The correlation with the energy minima of our solid curve is remarkable.
Even the even-odd phase changes (supershell structure) agree. This means
that the minima of the total energy determines the
most abundant island heights at the growing temperature which is
around the room temperature. According to Fig. \ref{sigma}
the energy differences between adjacent systems with even and odd
ML's of Pb are of the order of 0.05 eV per surface atom, {\em i.e.}
higher than the thermal energy. Menzel  {\em et al.} \cite{menzel}
estimated the evolution of a seven-layer island from a five-layer
island on Si(111) to occur through a barrier of 0.32 eV. They concluded
that the thermodynamics determines the most abundant island heights
but the kinetics is important in determining the height distribution.

\begin{figure}
\includegraphics[width=.8\columnwidth]{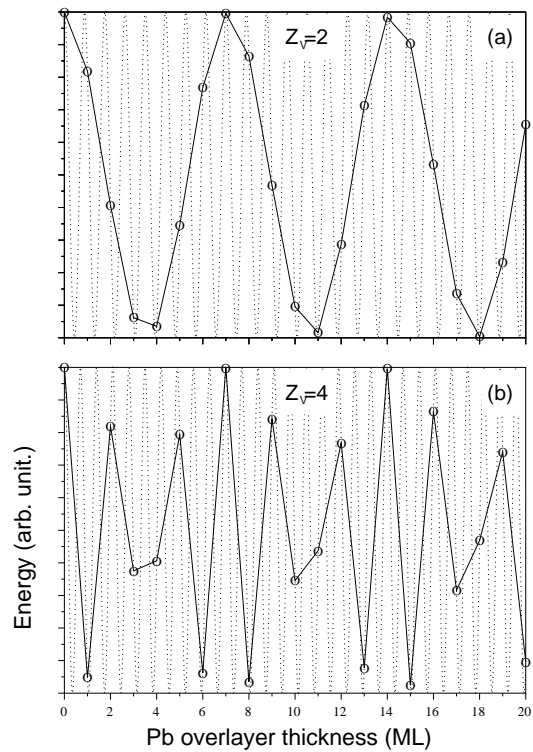}
\caption{\label{sigmateo}  Oscillating part of the total energy
calculated using Eqs. (2) and (4) for fcc metals grown in the (111) direction. 
$(a)$ and $(b)$ correspond to two and four valence electrons 
per atom, respectively. The dotted lines are the energies as continuous 
functions of the slab thickness and the solid lines connect the energies 
corresponding to completed ML's. }
\end{figure}

However, there is some mismatch between the theory and experiment in
Fig. \ref{sigma}. In experiments, islands of 13 ML's of Pb are
not especially abundant. The experimental situation in which the
substrate is not flat but contains steps or terraces may affect
the height distribution at this point \cite{private_rodolfo}. At the second beat our
model predicts similar energies for the island heights of 17 and 18 ML.
But according to experiments the 17 ML island should be more abundant.
The disagreement may be caused by a small error in the phase of
the energy oscillations in our model. In the beat regions the
relative energies are very sensitive on this phase. 

Hinch {\it et al.} \cite{hinch} have measured the shell and supershell 
structure of the Pb island heights with the HAS technique. During the
Pb deposition, a high intensity of the reflected He beam indicates smooth 
surfaces of islands of completed ML's. The heights of these islands are obtained indirectly from 
the deposition rate and the exposition time. The experiment shows the 
even-odd staggering which has an opposite phase in the comparison
with the STM results by Otero {\it et al.}  \cite{otero} and our calculations.
The beat positions by Hinch {\it et al.} are located at coverages of 10-11 
and 20-22 ML, which are slightly larger than those by Otero {\it et al.}
and our predictions. Thus, it seems evident that the indirect 
measurement overestimates the Pb overlayer thickness by one ML.

We have calculated the total energy of free-standing Pb-SJ
slabs. The results show oscillations with the same wavelength and a similar
amplitude as the dashed curve in Fig. \ref{sigma}. But the oscillating pattern
is now shifted about $0.8 a_0 \approx 1/7$ Pb ML forward. This 
changes the shell and supershell structure 
so that energy minima are shifted and the beat positions occur at 6-7 and 14-15 ML's, i.e., in a clear disagreement with the measurements.
We have compared our unsupported Pb-SJ results with 
those obtained for unsupported Pb films from pseudopotential calculations.
\cite{materzanini,wei} These more sophisticated evaluations give
similar wrong position of the first beat as our unsupported Pb-SJ slabs. 
The damping of the energy oscillations makes the second beat 
hardly recognizable in these calculations.
In conclusion, the energy oscillations are very sensitive to the correct determination of 
the potential barriers and it is crucial to take the Cu(111) substrate into account 
in the electronic structure calculations.

The shell structure of simple metal overlayers is a result of the periodic
sinking of new QWS's below the Fermi level when the thickness
of the overlayer increases. New QWS's increase the density 
of states (DOS) at the Fermi level giving the
oscillations in the energy with the wavelength of $\lambda_F/2$. 
Neglecting the damping we can
write the oscillating part of the total energy as a function of 
the thickness $D$ of the slab as $E_{osc}(z)= \cos \, (2\pi \nu D + \theta)$,
where $\nu=2/\lambda_F$ and $\theta$ is a phase that shifts the 
energy to the correct position depending on both the Pb-vacuum  
and the Cu-Pb interface properties. The energy values 
corresponding to $N$ completed ML's 
are obtained from
\begin{equation} \label{e_osc}
E_{osc}(N)= \cos \, (2\pi \nu d \, N + \theta).
\end{equation}
For a fcc metal grown in the (111) direction the interlayer spacing is $d=a/\sqrt{3}$, 
where $a$ is the lattice parameter. The Fermi wavelength is
$\lambda_F=(\frac{2\pi}{Z_v})^{1/3}a$, where $Z_v$ is the number of valence
electrons. These relations give
\begin{equation} \label{nud}
\nu \, d = \frac{1}{\sqrt{3}}\left( \frac{12 Z_v}{\pi} \right)^{1/3}\approx
0.903 \ Z_v^\frac{1}{3},
\end{equation}
and substituting this into Eq. \ref{e_osc} gives a general formula for $E_{osc}$ 
of a fcc metal grown in the (111) direction.

Using Eqs. (\ref{e_osc}) and (\ref{nud}) and setting $\theta=0$ we have
plotted in Fig. \ref{sigmateo} the oscillating part of the energy as a function of
the number of ML's for $Z_v$ = 2 and 4. The experimental
counterparts of these curves are the abundance spectra of islands
heights. For monovalent and divalent overlayer metals the
model predicts strong peaks occurring regularly after a certain number
of ML's. However, this pattern may be suppressed in the real
experiments, because the energy differences between neighboring
completed ML's are small and because the growing of a metastable 
island from the previous one requires several new ML's. For 
tri- and tetravalent metals the energy oscillates more rapidly. 
For $Z_v$ = 4, $\nu d$ is close to a half integer and the supershell 
structure with beats is clear. Note that the increase of electron
density as a function of $Z_v$ decreases the wavelength of $E_{osc}(z)$
as shown by the dashed curves in Fig. \ref{sigmateo}. This makes
the determination of the stable island heights very sensitive
to the variations of parameters such as the interplane distances
determining the overlayer thickness or the penetration of the
QWS wavefunctions into vacuum or/and into the Cu substrate.

In summary, we have analysed the electronic structures and stability 
of Pb islands grown on the Cu(111) surface as a function of the 
island height. The Cu(111) substrate is described by a new 1D 
pseudopotential and the Pb overlayer by the stabilized jellium model. 
As a function of Pb completed ML's our model gives quantum
well states in a good agreement with measurements. \cite{otero2} 
The total energy shows modulated odd-even
oscillations resembling the supershell structure of simple-metal 
atomic clusters and nanowires. The pattern correlates well with
the height abundance spectrum measured by Otero {\em et al.}
\cite{otero}. We demonstrate that a proper modeling of the 
Cu(111) substrate plays a crucial role in predicting the beat
positions of the abundance spectra.

We thank R. Miranda for useful discussions. This work was partially supported 
by the University of the Basque Country, Departamento de Educaci\'{o}n del 
Gobierno Vasco, Spanish MCyT (MAT 2001-0946, PB98-0870-C02 and MAT 2002-04087-C02-01) and by the Academy of Finland through its Centre of Excellence Program (2000-2005). 


\bibliography{bibliografia}

\begin{thebibliography}{21}
\expandafter\ifx\csname natexlab\endcsname\relax\def\natexlab#1{#1}\fi
\expandafter\ifx\csname bibnamefont\endcsname\relax
  \def\bibnamefont#1{#1}\fi
\expandafter\ifx\csname bibfnamefont\endcsname\relax
  \def\bibfnamefont#1{#1}\fi
\expandafter\ifx\csname citenamefont\endcsname\relax
  \def\citenamefont#1{#1}\fi
\expandafter\ifx\csname url\endcsname\relax
  \def\url#1{\texttt{#1}}\fi
\expandafter\ifx\csname urlprefix\endcsname\relax\def\urlprefix{URL }\fi
\providecommand{\bibinfo}[2]{#2}
\providecommand{\eprint}[2][]{\url{#2}}

\bibitem[{\citenamefont{Knight et~al.}(1984)\citenamefont{Knight, Clemenger,
  de~Heer, Saunders, Chou, and Cohen}}]{knight84}
\bibinfo{author}{\bibfnamefont{W.}~\bibnamefont{Knight}},
  \bibinfo{author}{\bibfnamefont{K.}~\bibnamefont{Clemenger}},
  \bibinfo{author}{\bibfnamefont{W.}~\bibnamefont{de~Heer}},
  \bibinfo{author}{\bibfnamefont{W.~A.} \bibnamefont{Saunders}},
  \bibinfo{author}{\bibfnamefont{M.}~\bibnamefont{Chou}}, \bibnamefont{and}
  \bibinfo{author}{\bibfnamefont{M.~L.} \bibnamefont{Cohen}},
  \bibinfo{journal}{Phys. Rev. Lett.} \textbf{\bibinfo{volume}{52}},
  \bibinfo{pages}{2141} (\bibinfo{year}{1984}).

\bibitem[{\citenamefont{Yanson et~al.}(1999)\citenamefont{Yanson, Yanson, and
  van Ruitenbeek}}]{yanson2}
\bibinfo{author}{\bibfnamefont{A.}~\bibnamefont{Yanson}},
  \bibinfo{author}{\bibfnamefont{I.}~\bibnamefont{Yanson}}, \bibnamefont{and}
  \bibinfo{author}{\bibfnamefont{J.}~\bibnamefont{van Ruitenbeek}},
  \bibinfo{journal}{Nature} \textbf{\bibinfo{volume}{400}},
  \bibinfo{pages}{144} (\bibinfo{year}{1999}).

\bibitem[{\citenamefont{Hinch et~al.}(1989)\citenamefont{Hinch, Koziol,
  Toennies, and Zhang}}]{hinch}
\bibinfo{author}{\bibfnamefont{B.}~\bibnamefont{Hinch}},
  \bibinfo{author}{\bibfnamefont{C.}~\bibnamefont{Koziol}},
  \bibinfo{author}{\bibfnamefont{J.}~\bibnamefont{Toennies}}, \bibnamefont{and}
  \bibinfo{author}{\bibfnamefont{G.}~\bibnamefont{Zhang}},
  \bibinfo{journal}{Europhys. Lett.} \textbf{\bibinfo{volume}{10}},
  \bibinfo{pages}{341} (\bibinfo{year}{1989}).

\bibitem[{\citenamefont{Crommie et~al.}(1993)\citenamefont{Crommie, Lutz, and
  Eigler}}]{crommie}
\bibinfo{author}{\bibfnamefont{M.~F.} \bibnamefont{Crommie}},
  \bibinfo{author}{\bibfnamefont{C.~P.} \bibnamefont{Lutz}}, \bibnamefont{and}
  \bibinfo{author}{\bibfnamefont{D.~M.} \bibnamefont{Eigler}},
  \bibinfo{journal}{Science} \textbf{\bibinfo{volume}{262}},
  \bibinfo{pages}{218} (\bibinfo{year}{1993}).

\bibitem[{\citenamefont{Otero et~al.}(2002)\citenamefont{Otero, de~Parga, and
  Miranda}}]{otero}
\bibinfo{author}{\bibfnamefont{R.}~\bibnamefont{Otero}},
  \bibinfo{author}{\bibfnamefont{A.~V.} \bibnamefont{de~Parga}},
  \bibnamefont{and} \bibinfo{author}{\bibfnamefont{R.}~\bibnamefont{Miranda}},
  \bibinfo{journal}{Phys.\ Rev.\ B} \textbf{\bibinfo{volume}{66}},
  \bibinfo{pages}{115401} (\bibinfo{year}{2002}).

\bibitem[{\citenamefont{Schulte}(1976)}]{schulte}
\bibinfo{author}{\bibfnamefont{F.}~\bibnamefont{Schulte}},
  \bibinfo{journal}{Surf. Sci.} \textbf{\bibinfo{volume}{55}},
  \bibinfo{pages}{427} (\bibinfo{year}{1976}).

\bibitem[{\citenamefont{Feibelman}(1983)}]{feibelman}
\bibinfo{author}{\bibfnamefont{P.}~\bibnamefont{Feibelman}},
  \bibinfo{journal}{Phys.\ Rev.\ B.} \textbf{\bibinfo{volume}{27}},
  \bibinfo{pages}{1991} (\bibinfo{year}{1983}).

\bibitem[{\citenamefont{Sarr\'ia et~al.}(2000)\citenamefont{Sarr\'ia,
  Henriques, Fiolhais, and Pitarke}}]{sarria}
\bibinfo{author}{\bibfnamefont{I.}~\bibnamefont{Sarr\'ia}},
  \bibinfo{author}{\bibfnamefont{C.}~\bibnamefont{Henriques}},
  \bibinfo{author}{\bibfnamefont{C.}~\bibnamefont{Fiolhais}}, \bibnamefont{and}
  \bibinfo{author}{\bibfnamefont{J.}~\bibnamefont{Pitarke}},
  \bibinfo{journal}{Phys. Rev. B.} \textbf{\bibinfo{volume}{62}},
  \bibinfo{pages}{1699} (\bibinfo{year}{2000}).

\bibitem[{\citenamefont{Saalfrank}(1992)}]{saalfrank}
\bibinfo{author}{\bibfnamefont{P.}~\bibnamefont{Saalfrank}},
  \bibinfo{journal}{Surf. Sci.} \textbf{\bibinfo{volume}{274}},
  \bibinfo{pages}{449} (\bibinfo{year}{1992}).

\bibitem[{\citenamefont{Materzanini et~al.}(2001)\citenamefont{Materzanini,
  Saalfrank, and Lindan}}]{materzanini}
\bibinfo{author}{\bibfnamefont{G.}~\bibnamefont{Materzanini}},
  \bibinfo{author}{\bibfnamefont{P.}~\bibnamefont{Saalfrank}},
  \bibnamefont{and} \bibinfo{author}{\bibfnamefont{P.}~\bibnamefont{Lindan}},
  \bibinfo{journal}{Phys.\ Rev.\ B} \textbf{\bibinfo{volume}{63}},
  \bibinfo{pages}{235405} (\bibinfo{year}{2001}).

\bibitem[{\citenamefont{Wei and Chou}(2002)}]{wei}
\bibinfo{author}{\bibfnamefont{C.}~\bibnamefont{Wei}} \bibnamefont{and}
  \bibinfo{author}{\bibfnamefont{M.}~\bibnamefont{Chou}},
  \bibinfo{journal}{Phys.\ Rev.\ B} \textbf{\bibinfo{volume}{66}},
  \bibinfo{pages}{233408} (\bibinfo{year}{2002}).

\bibitem[{\citenamefont{Kiejna et~al.}(1999)\citenamefont{Kiejna, Peisert, and
  P.Scharoch}}]{kiejna}
\bibinfo{author}{\bibfnamefont{A.}~\bibnamefont{Kiejna}},
  \bibinfo{author}{\bibfnamefont{J.}~\bibnamefont{Peisert}}, \bibnamefont{and}
  \bibinfo{author}{\bibnamefont{P.Scharoch}}, \bibinfo{journal}{Surf. Sci.}
  \textbf{\bibinfo{volume}{432}}, \bibinfo{pages}{54} (\bibinfo{year}{1999}).

\bibitem[{\citenamefont{Perdew et~al.}(1990)\citenamefont{Perdew, Tran, and
  Smith}}]{perdew}
\bibinfo{author}{\bibfnamefont{J.}~\bibnamefont{Perdew}},
  \bibinfo{author}{\bibfnamefont{H.}~\bibnamefont{Tran}}, \bibnamefont{and}
  \bibinfo{author}{\bibfnamefont{E.}~\bibnamefont{Smith}},
  \bibinfo{journal}{Phys.\ Rev.\ B} \textbf{\bibinfo{volume}{42}},
  \bibinfo{pages}{11627} (\bibinfo{year}{1990}).

\bibitem[{\citenamefont{Ashcroft and Mermin}(1981)}]{ashcroft}
\bibinfo{author}{\bibfnamefont{N.}~\bibnamefont{Ashcroft}} \bibnamefont{and}
  \bibinfo{author}{\bibfnamefont{N.}~\bibnamefont{Mermin}},
  \emph{\bibinfo{title}{Solid State Physics}}
  (\bibinfo{publisher}{Hold-Saunders International Editions},
  \bibinfo{year}{1981}).

\bibitem[{\citenamefont{Chulkov et~al.}(1999)\citenamefont{Chulkov, Silkin, and
  Echenique}}]{chulkov}
\bibinfo{author}{\bibfnamefont{E.}~\bibnamefont{Chulkov}},
  \bibinfo{author}{\bibfnamefont{V.}~\bibnamefont{Silkin}}, \bibnamefont{and}
  \bibinfo{author}{\bibfnamefont{P.}~\bibnamefont{Echenique}},
  \bibinfo{journal}{Surf. Sci.} \textbf{\bibinfo{volume}{437}},
  \bibinfo{pages}{330} (\bibinfo{year}{1999}).

\bibitem[{\citenamefont{Ogando et~al.}(2003)\citenamefont{Ogando, Zabala,
  Chulkov, and Puska}}]{ogando}
\bibinfo{author}{\bibfnamefont{E.}~\bibnamefont{Ogando}},
  \bibinfo{author}{\bibfnamefont{N.}~\bibnamefont{Zabala}},
  \bibinfo{author}{\bibfnamefont{E.}~\bibnamefont{Chulkov}}, \bibnamefont{and}
  \bibinfo{author}{\bibfnamefont{M.}~\bibnamefont{Puska}}
  (\bibinfo{year}{2003}), \bibinfo{note}{to be submitted}.

\bibitem[{\citenamefont{Otero et~al.}(2000)\citenamefont{Otero, de~Parga, and
  Miranda}}]{otero2}
\bibinfo{author}{\bibfnamefont{R.}~\bibnamefont{Otero}},
  \bibinfo{author}{\bibfnamefont{A.~V.} \bibnamefont{de~Parga}},
  \bibnamefont{and} \bibinfo{author}{\bibfnamefont{R.}~\bibnamefont{Miranda}},
  \bibinfo{journal}{Surf. Sci.} \textbf{\bibinfo{volume}{447}},
  \bibinfo{pages}{143} (\bibinfo{year}{2000}).

\bibitem[{\citenamefont{de~Heer}(1993)}]{heer}
\bibinfo{author}{\bibfnamefont{W.}~\bibnamefont{de~Heer}},
  \bibinfo{journal}{Rev.\ Mod.\ Phys.} \textbf{\bibinfo{volume}{65}},
  \bibinfo{pages}{611} (\bibinfo{year}{1993}).

\bibitem[{\citenamefont{Yanson et~al.}(2000)\citenamefont{Yanson, Yanson, and
  van Ruitenbeek}}]{yanson1}
\bibinfo{author}{\bibfnamefont{A.}~\bibnamefont{Yanson}},
  \bibinfo{author}{\bibfnamefont{I.}~\bibnamefont{Yanson}}, \bibnamefont{and}
  \bibinfo{author}{\bibfnamefont{J.}~\bibnamefont{van Ruitenbeek}},
  \bibinfo{journal}{Phys.\ Rev.\ Lett.} \textbf{\bibinfo{volume}{84}},
  \bibinfo{pages}{5832} (\bibinfo{year}{2000}).

\bibitem[{\citenamefont{Menzel et~al.}(2003)\citenamefont{Menzel, Kammler,
  Conrad, Yeh, Hupalo, and Tringides}}]{menzel}
\bibinfo{author}{\bibfnamefont{A.}~\bibnamefont{Menzel}},
  \bibinfo{author}{\bibfnamefont{M.}~\bibnamefont{Kammler}},
  \bibinfo{author}{\bibfnamefont{E.}~\bibnamefont{Conrad}},
  \bibinfo{author}{\bibfnamefont{V.}~\bibnamefont{Yeh}},
  \bibinfo{author}{\bibfnamefont{M.}~\bibnamefont{Hupalo}}, \bibnamefont{and}
  \bibinfo{author}{\bibfnamefont{M.}~\bibnamefont{Tringides}},
  \bibinfo{journal}{Phys. Rev. B.} \textbf{\bibinfo{volume}{67}},
  \bibinfo{pages}{165314} (\bibinfo{year}{2003}).

\bibitem[{\citenamefont{Miranda and de~Parga}(2003)}]{private_rodolfo}
\bibinfo{author}{\bibfnamefont{R.}~\bibnamefont{Miranda}} \bibnamefont{and}
  \bibinfo{author}{\bibfnamefont{A.~V.} \bibnamefont{de~Parga}}
  (\bibinfo{year}{2003}), \bibinfo{note}{private communication}.

\end{thebibliography}

\end{document}